\documentclass[letterpaper,12pt,fleqn]{article}
\pdfoutput=1
\usepackage{jheppub}
\usepackage{xspace}
\usepackage{engord}
\usepackage{xspace}
\usepackage{epstopdf}

\newcommand{\peff}{p_\mathrm{eff}}
\newcommand{\pvap}{p_\mathrm{v}}
\newcommand{\bulk}{\left.\frac{\zeta}{s}\right|_\mathrm{crit}}
\newcommand{\bulktext}{\left.\zeta/s\right|_\mathrm{crit}}
\newcommand{\zetas}{\frac{\zeta}{s}}
\newcommand{\zetastext}{\zeta/s}

\newcommand{\DU}[1]{\nabla_{#1} u^{#1}}

\newcommand{\cssqu}{c_\mathrm{s}^2}
\newcommand{\bulkv}{bulk viscosity\xspace}

\newcommand{\hic}{heavy-ion collisions\xspace}

\newcommand{\fst}{\engordnumber{1}\xspace}
\newcommand{\snd}{\engordnumber{2}\xspace}

\newcommand{\fmc}{\mathrm{fm/c}}

\newcommand{\beq}{\begin{eqnarray}}
\newcommand{\eeq}{\end{eqnarray}}

\newcommand{\thetitle}{Onset of cavitation in the quark--gluon plasma}
\title{\thetitle}
\author{Mathis Habich}
\author{Paul Romatschke}
\affiliation{Department of Physics\\ 390 UCB, University of Colorado, Boulder, CO 80309-0390, USA}
\emailAdd{mathis.habich@colorado.edu}
\emailAdd{paul.romatschke@colorado.edu}
\keywords{QGP, hydrodynamics, \hic, bulk viscosity}

\abstract{We study the onset of bubble formation (cavitation) in the quark--gluon plasma as a result of the reduction of the effective pressure from bulk-viscous corrections. By calculating velocity gradients in typical models for quark--gluon plasma evolution in heavy-ion collisions, we obtain results for the critical bulk viscosity  above which cavitation occurs. Since present experimental data for \hic seems inconsistent with the presence of bubbles above the phase transition temperature of QCD, our results may be interpreted as an upper limit of the bulk viscosity in nature. Our results indicate that bubble formation is consistent with the expectation of hadronisation in low-temperature QCD.  }

\begin{document}
\maketitle
\section{Introduction}
\label{sec:intro}
The experimental \hic programme conducted at the Relativistic Heavy Ion Collider and the Large Hadron Collider strongly suggests that the quark--gluon plasma (QGP) formed in these collisions behaves like an almost ideal fluid \cite{RomatschkeNewDevRelHydro,HeinzCollectiveFlowViscHIC,MuellerResultsFromPbPbLHC,SchaeferNearlyPerfectFluidityColdHot}.
This fluid is very well-described by relativistic hydrodynamics \cite{SchenkeEventbyEventAnisotropicFlowHICYM}. 
For an ordinary fluid such as water, the effective pressure  can be different than the equilibrium pressure and, in particular, in some situations it can drop below the vapour pressure. In this case, the thermodynamically preferred phase becomes the gas phase, and a vapour bubble forms inside the fluid, a phenomenon known as 'cavitation'\footnote{We note that if the bubble is unstable, then it quickly collapses. Nevertheless, the onset of cavitation signals an instability in the fluid evolution.}. Mainly, high fluid velocities trigger cavitation in liquids.  
In the case of relativistic fluids such as the QGP studied in \hic, 
cavitation would imply a phase transition from a deconfined plasma phase of quarks and gluons to a confined hadron-gas phase. The resulting medium would be highly inhomogeneous with (possibly short-lived) hadron gas bubbles expanding and collapsing in an otherwise laminar fluid. However, the apparent success of describing experimental data by relatively simple, laminar fluid flows seems inconsistent with the presence of hadron gas bubbles, or even the onset of fluid instabilities in the high-temperature QGP. In the present work, we will thus make the assumption that cavitation does \emph{not} occur in the experimentally observed QGP. We will study cavitation in relativistic fluids with a given bulk viscosity coefficient and then proceed to rule out bulk viscosity values under the above assumption.

In this article, the effective pressure is defined as one third of the trace of the pressure tensor \cite{BuchelCavitationEffectsConfineDeconfine}. For this definition, the shear viscous contribution
to the pressure cancels, whereas the bulk-viscous contribution remains present. This differs from other approaches taken in the literature, which focused on single components of the pressure tensor
\cite{RajagopalBulkViscandCavitation,KlimekCavitationHolographicsQGP,SreekanthCavitationThermalPhotonProductionHIC,SreekanthShearViscCavitationHydroLHC,SreekanthThermalPhotonsQGPnonidealeffects} where the shear viscous contribution is present and important. Because shear-viscous effects will add on to the effects considered in this work, cavitation could occur in regions which --- in our analysis --- are found to be stable, but not the other way around. One caveat of our approach is that 
in the calculations that follow, the bulk-viscous contributions to the fluid flow profiles themselves have been neglected for simplicity. In principle, these contributions should be taken into account, but in practice, one expects the corrections to be small as long as the bulk viscosity coefficient itself is small \cite{HeinzInterplayShearBulkViscGeneratingFlowHIC}. Thus, our approach essentially amounts to a linear-response treatment of bulk-viscous effects in fully non-linear, shear-viscous fluid dynamics.

This work is organised as follows: in section \ref{sec:definition}, 
cavitation for relativistic hydrodynamics is defined. In section \ref{sec:calcupper}, the main idea of constraining bulk viscosity is elucidated for \fst and \snd order hydrodynamics. Section \ref{sec:HIC} applies this framework to \hic, i.e., the critical bulk viscosity for cavitation is calculated for analytical and numerical flow profiles, as well as different equations of state. We present our conclusions and an upper limit for the QCD bulk viscosity in section \ref{sec:conc}.

\section{Bulk-viscous bubble formation in relativistic hydrodynamics}
\label{sec:definition}

Cavitation can be defined as the drop of pressure below the saturated vapour pressure of the particular liquid (see p. 6 in ref. \cite{brennen1995cavitation}). This definition needs to be revisited for relativistic fluids 
which can have a pressure tensor that differs strongly from equilibrium. Formally, starting from a standard decomposition of  the energy-momentum tensor
\beq
T^{\mu \nu} 
&=&
\epsilon u^\mu u^\nu + \left(p-\Pi\right) \Delta^{\mu \nu} + \pi^{\mu \nu}\,,
\eeq
with $\epsilon, p, u^\mu$ the energy density, pressure and fluid four velocity, and the projector $\Delta^{\mu\nu}=g^{\mu\nu}-u^\mu u^\nu$ with a mostly minus signature metric tensor $g^{\mu\nu}$, we identify $\Pi,\pi^{\mu\nu}$ as the bulk- and shear-viscous stress tensor components.
In this work, we concentrate on the effective, local pressure defined in three dimensions as 
\beq
\peff 
&\equiv& 
-\frac{1}{3} \Delta_{\mu\nu} T^{\mu\nu}= p - \Pi \,.
\label{eq:peff}
\eeq
This preserves the normal, intuitive definition of pressure in the rest frame and, being a scalar, is easy to interpret in non-equilibrium situation.\\
Akin to ref. \cite{brennen1995cavitation}, the \emph{occurrence of cavitation} can be mathematically defined as
\beq
\peff &<& \pvap\,, \label{eq:cav}
\eeq
with $\pvap$ the 'vapour' pressure of a different thermodynamic phase having the same energy density.
In words, this means that if the effective pressure $\peff$ of a QGP falls below $\pvap$, 
then the liquid will undergo a phase transition to the hadron-gas phase (often referred to as 'freeze-out' in the language of \hic) and a (small) gas bubble will form.
The definition (\ref{eq:cav}) is intuitively much easier to interpret than the case considered by most other authors \cite{RajagopalBulkViscandCavitation,KlimekCavitationHolographicsQGP,SreekanthCavitationThermalPhotonProductionHIC,SreekanthShearViscCavitationHydroLHC,SreekanthThermalPhotonsQGPnonidealeffects},
where a single component of the pressure tensor drops below the vapour pressure, whereas other component(s) will generally be larger than the vapour pressure.

\subsection{Critical bulk viscosity in first and second order hydrodynamics}
\label{sec:calcupper}

After establishing a criterion for cavitation, the critical bulk viscosity for the onset of cavitation can be calculated by assuming the validity of hydrodynamics.

\paragraph{For \texorpdfstring{\engordnumber{1}}{first} order hydrodynamics}
The effective pressure \eqref{eq:peff} up to \fst order gradients \cite{RomatschkeNonequilibrium} for a non-conformal fluid is
\beq
\peff &=& p-\Pi \approx p - \zeta \DU{\mu}\,.
\label{eq:1stpeff}
\eeq
The critical bulk viscosity is defined as the maximum value of $\zeta$ for which the fluid flow is still \emph{non-cavitating}. Assuming the sign of the gradient $\DU{\mu}$ to be positive (this is the case for all the scenarios we consider below), one finds
\beq
\bulk \equiv \frac{\left(p-\pvap \right)T}{\left( \epsilon +p \right)\nabla_\mu u^\mu}  \,,
\label{eq:1stzeta}
\eeq
where the bulk viscosity was divided by the entropy density $s=(\epsilon +p)/T$, yielding a dimensionless ratio. 

\paragraph{For \texorpdfstring{\snd}{second} order hydrodynamics}
\label{sec:2ndzeta}

In order to assess the accuracy of \fst order calculation, one can consider the effect of \snd order gradients. Expanding the viscosity scalar $\Pi$ of eq. \eqref{eq:peff} up to \snd order in gradients yields for a flat space--time \cite{RomatschkeNonequilibrium}
\beq
\peff 
=
 p - \zeta \DU{\mu} + \zeta\tau_\Pi \mathrm{D} \left( \DU{\mu} \right) +\xi_1\sigma^{\mu \nu} \sigma_{\mu \nu} + \xi_2 \left( \DU{\mu} \right)^2  \label{eq:2nd}\,.
\label{eq:2ndzetageneral}
\eeq
Most of the \snd order transport coefficients $\tau_\Pi,\xi_1,\xi_2$ are poorly known for most quantum field theories. However, in a particular strong coupling\footnote{Note that by construction, this particular theory  only includes terms of linear order in bulk viscosity, e.g., terms of the form $\zeta^2$ are absent in the transport coefficients.} (see ref. \cite{KanitscheiderUniversalhydrodynamicsnon-confbranes}),  these have been calculated \cite{RomatschkeNonequilibrium}:
\beq
\eta  
&=& 
\frac{3\zeta}{2(1-3\cssqu)}\,, \\
\zeta \tau_\Pi 
&=&
\zeta \tau_\pi = \frac{\zeta}{\epsilon + p} \eta (4-\ln 4) \label{eq:taupi}\,,\\
\xi_1 
&=& 
\frac{\lambda_1}{3}\left( 1-3\cssqu \right) 
= 
\frac{2\eta^2}{3(\epsilon+p)}\left( 1-3\cssqu \right) 
=
\frac{\zeta}{\epsilon+p} \eta\,,\\
\xi_2 
&=&
\frac{2\eta\tau_\Pi \cssqu}{3}\left( 1-3\cssqu \right)
=
\frac{\zeta\tau_\Pi \cssqu}{1-3\cssqu} 
=
\frac{\zeta}{\epsilon +p} \eta\cssqu (4-\ln4)\,.
\eeq
By expressing these transport coefficients in terms of the speed of sound squared $\cssqu$, $\zeta$, and shear viscosity $\eta$, one finds the critical bulk viscosity in \snd order hydrodynamics:
\beq
\bulk \equiv 
\frac{\left(\pvap -p \right)T}{\left( \epsilon +p \right)} \left[\DU{\mu} - \frac{\eta}{s} \frac{(4-\ln4)\left(\mathrm{D}\DU{\mu}+c_s^2 \left(\DU{\mu} \right)^2\right)
+\sigma^{\mu \nu} \sigma_{\mu \nu}}{T}\right]^{-1}\,. \nonumber \\
\label{eq:2ndzeta}
\eeq
Note that in the \snd order result, there is a pole in the critical bulk viscosity once the \snd order gradient terms become as large as the \fst order terms. We find that for QCD this typically happens at very low temperatures (far below $T_c$), where one does not expect a hydrodynamic description to be applicable in the first place.
In weak coupling, the transport coefficients typically lead to a quadratic dependence of $\zetastext$ \cite{JaiswalPartProdHIChydroApproach,JaiswalRel2nddissipativeHydrofromEntropy} which is beyond the scope of our linear-response treatment.\\
To recapitulate, this method assumes a conformal, hydrodynamic description and uses the resulting flow profile to constrain the maximum value of $\zetastext$ by requiring that the \emph{effective pressure does not drop below zero}.

\section{Cavitation in \hic}
\label{sec:HIC}

In this section, the critical bulk viscosity coefficient for the onset of cavitation is calculated for specific flow profiles used in the modelling of \hic. Specifically, the flow gradients are calculated for Bjorken flow \cite{BjorkenHighlyRelNNcollisions}; Gubser flow \cite{GubserSymmetryConstraintsGeneralizationsBjorken,GubserConfHydrodynamicsMinkowskidS};  and a numerical solver for relativistic, viscous hydrodynamics in 2+1 dimensions \cite{RomatschkeViscInfPerfect,BaierCausalHydroForHIC}. All these flow profiles are for \emph{conformal fluids}, e.g., they ignore effects of bulk viscosity in the flow itself (see the discussion in section \ref{sec:intro}). 
In this entire chapter, the vapour pressure $\pvap$ is chosen to be zero:
\beq
\pvap &\equiv& 0\,,
\eeq
which will result in the most conservative estimates of cavitation since higher values of $\pvap$ would decrease $\bulktext$ (e.g., see eq. \eqref{eq:1stzeta}).

\subsection{Bjorken flow}
\label{sec:ieos}

The set-up that was suggested by Bjorken \cite{BjorkenHighlyRelNNcollisions} in 1982 can be utilised to extract a benchmark value on the \bulkv. It is particularly simple to use Milne coordinates $x^\mu=(\tau,x,y,\eta)$ because the fluid velocity becomes $u^\mu = (1,0,0,0)^T$, i.e., the fluid is locally at rest. The velocity gradient simply depends on the Christoffel symbols for Milne coordinates
\beq
\DU{\mu} &=& \frac{1}{\tau}\,,
\eeq
whereas the temperature evolution is governed by
\beq
T &=& T_0 \left(\tau/\tau_0 \right)^{-\cssqu}\,.
\eeq
This framework is comparatively simple because it entirely neglects transverse, spatial dynamics of the QGP. For an ideal equation of state the dynamics is completely governed by the gradients: evidently, the gradient is always positive for $\tau>0$ for \fst order hydrodynamics; however, for \snd order gradients, the gradients become negative after diverging for early times and/or low temperatures ($\tau T \ll 1$).

\begin{figure}[ht]
\input{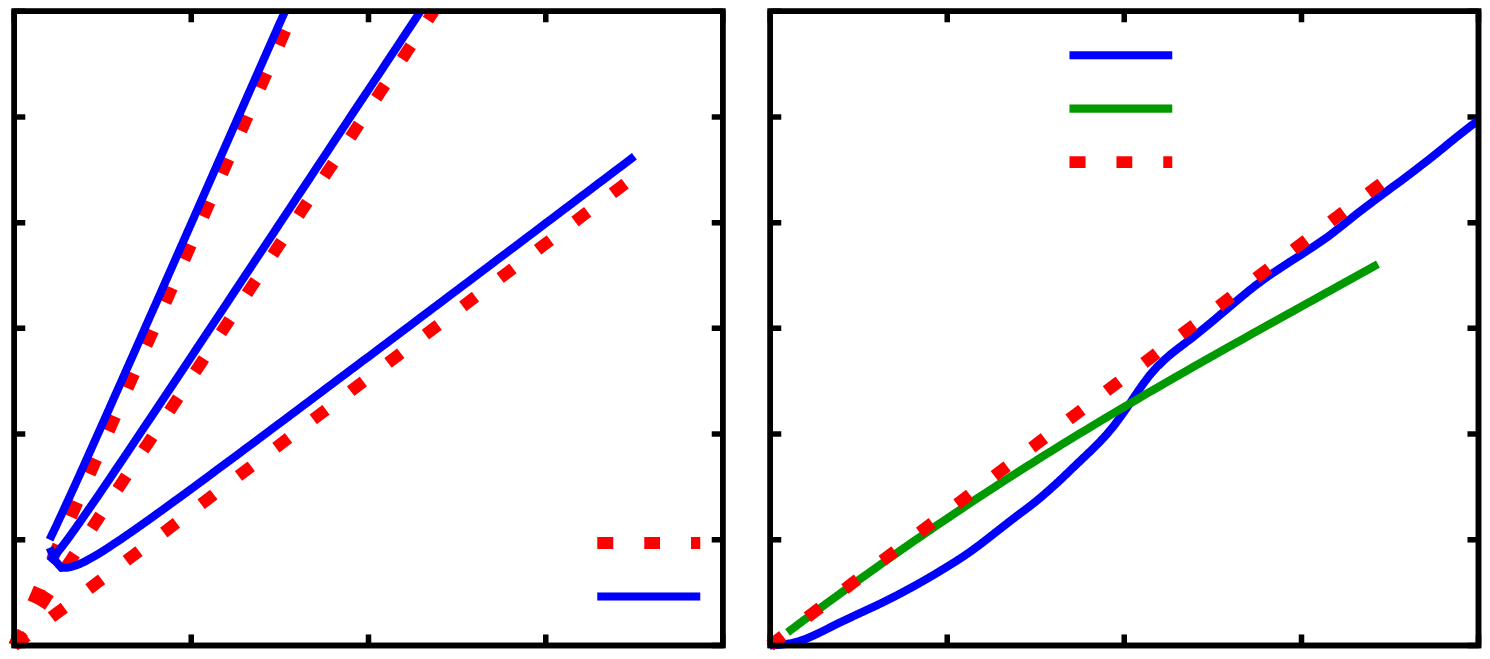}
\caption{Critical bulk viscosity as a function of temperature for ideal equation of state. Areas above respective lines of $\bulktext$ are regions where cavitation occurs.
Left: comparison of \fst and \snd order results for Bjorken flow. \hspace{\textwidth}
Right: comparison between Bjorken flow, Gubser flow, and numerical computations.}
\label{fig:I_bjorken}
\end{figure}

The critical bulk viscosity for Bjorken flow is monotonously increasing with temperature. By taking \snd order gradients into account, only small changes are present, as can be seen in figure \ref{fig:I_bjorken}. The effect of these higher-order terms is small due to the small numerical values of the \snd order transport coefficients.

\subsection{Gubser flow}
By expressing the flow profile that was proposed by Gubser \cite{GubserSymmetryConstraintsGeneralizationsBjorken} in Milne coordinates, one finds a fluid flow profile
\beq
u^\mu = \left(\frac{1+q^2 r^2 +q^2 \tau^2}{2q\tau\sqrt{1+g^2}},\,\frac{q r}{\sqrt{1+g^2}},\,0,\,0 \right)^\mathrm{T}
\label{eq:gubserflow}
\eeq
and a temperature profile
\beq
T 
&=& 
\frac{1}{\tau f_*^{1/4}} 
\left\{ 
\frac{\hat{T}_0}{\left(1+g^2\right)^{1/3}} + \frac{H_0}{\sqrt{1+g^2}}
\left[
1-(1+g^2)^{1/6}\, {_2F_1}\left(\frac{1}{2},\frac{1}{6},\frac{3}{2},-g^2 \right) 
\right] 
\right\},
\eeq
where
\beq
g &=& \frac{1+q^2 r^2 -q^2 \tau^2}{2q\tau}\,,\quad f_*^{1/4} \,, \quad \hat{T}=5.55\,, \quad H_0=0.33 \,, \quad  1/q = 4.3 \mathrm{fm} \nonumber\,.
\eeq
The variables $\tau$ and $r$ denote the proper time and radial distance, respectively. The specific form of fluid velocity and temperature arises from symmetry considerations \cite{GubserSymmetryConstraintsGeneralizationsBjorken,GubserConfHydrodynamicsMinkowskidS}. 
In comparison to Bjorken flow, this flow profile is more realistic because radial velocities are non-vanishing; hence, the transverse dynamics is not neglected but fixed to have a unique, analytical form. By comparing the different orders of Gubser flow, one sees a similar behaviour to Bjorken flow, i.e., higher-order gradients decrease the denominator; thus, \snd order terms increase the value of $\bulktext$
(see figure \ref{fig:I_bjorken}). 

\subsection{Numerical results}

The last flow profile stems from a numerical simulation that fully includes transverse dynamics ("VH2+1", see refs. \cite{BaierCausalHydroForHIC,RomatschkeViscInfPerfect} for details) for an initial condition of a central $Au+Au$ collision at $\sqrt{s}=200$ GeV per nucleon pair. 
It was initialised with vanishing flow at early times, such that the high-temperature $\bulktext$ behaviour matches the Bjorken flow result, as it should. For late times, the significant fluid velocity gradients differ from both the Bjorken and Gubser flow results, resulting in a different $\bulktext$ behaviour at low temperatures (see figure \ref{fig:I_bjorken}).

\subsection{QCD equation of state}
\label{sec:qcdeos}

\begin{figure}[ht]
\input{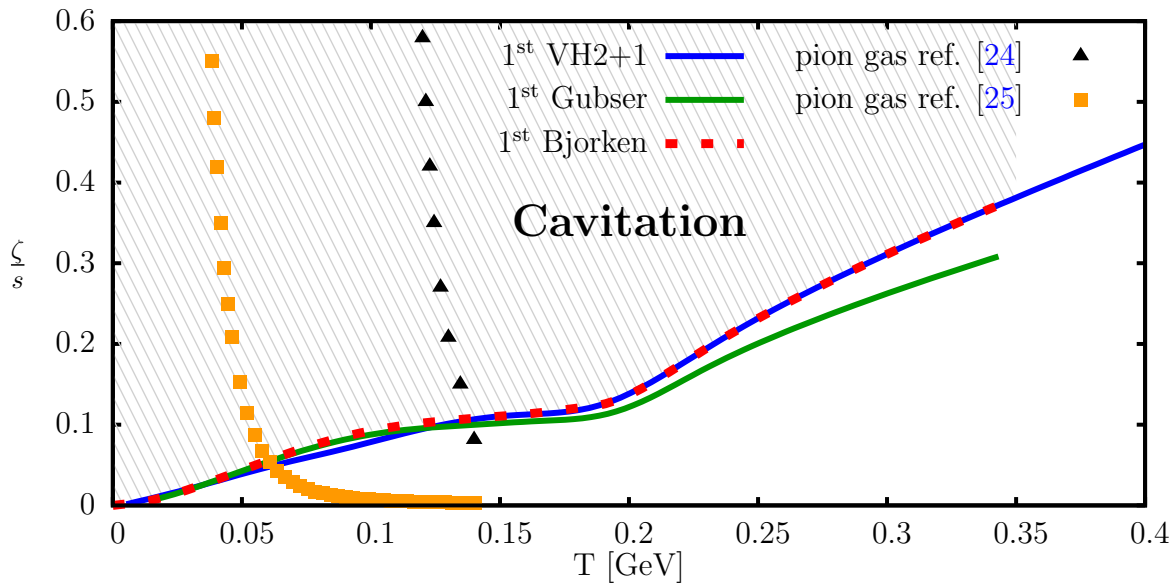}
\caption{Bulk viscosity over entropy density ratio as a function of temperature. Shown are results for the lowest critical bulk viscosity coefficient $\bulktext$ using different flow profiles and a QCD equation of state. For higher viscosity values, we predict bubbles to form in the liquid ('cavitation'). For comparison we also show the result of a calculation of $\zeta/s$ for two pion gases from refs. \cite{MooreBulkViscPionGas,TorresBulkViscoflowTstronglyinteractingmatter}. The pion gas is calculated up to $T=140$ MeV.}
\label{fig:Qall}
\end{figure}

For a realistic model of the QGP, we have repeated the above calculations for $\bulktext$ with a QCD equation of state (see ref. \cite{LaineQuarkMassThresholdsInQCDThermodynamics}). In figure \ref{fig:Qall}, the \emph{lowest} value of $\bulktext$ for Bjorken, Gubser and numerical flow profiles, respectively, is shown. For comparison, we also show the result of calculations of $\zeta/s$ for a pion gas from refs. \cite{MooreBulkViscPionGas,TorresBulkViscoflowTstronglyinteractingmatter}. The former is performed at chemical equilibrium; whereas, the latter being an out-of-chemical equilibrium calculation for which elastic scattering is the dominant process \cite{TorresPersonalCom,TorresBulkViscoflowTstronglyinteractingmatter}. In dynamical \hic $\zetastext$ is most likely to lie between these curves.
At very high temperature, one could also compare to perturbative QCD calculations from ref. \cite{ArnoldBulkViscHiTQCD} where $\zeta/s\sim 0.01 \alpha_s^2$, which tends to fall as a function of temperature, whereas $\bulktext$ rises with temperature. 
Thus, we presume that cavitation is a phenomenon of low temperatures --- not of high temperatures.

\newcommand{\pigasfromrefmoore}{pion gas ref. \cite{MooreBulkViscPionGas}}
\newcommand{\pigasfromreftorres}{pion gas ref. \cite{TorresBulkViscoflowTstronglyinteractingmatter}}

\section{Conclusion}
\label{sec:conc}

In this work, we have studied the onset of bubble formation (cavitation) in the QGP resulting from the presence of bulk-viscous terms in relativistic hydrodynamics.  We found that at temperatures $T<140$ MeV, a bulk viscosity coefficient smaller than that expected from a pion gas leads to the formation of hadron gas bubbles in the QGP liquid (see also refs. \cite{TorresBulkViscoflowTstronglyinteractingmatter,MooreBulkViscPionGas,Torrieri:2007fb,Torrieri:2008ip,RajagopalBulkViscandCavitation,BuchelCavitationEffectsConfineDeconfine,SreekanthCavitationThermalPhotonProductionHIC,SreekanthShearViscCavitationHydroLHC,KlimekCavitationHolographicsQGP}). This may be interpreted as the known freeze-out phenomenon in \hic where the plasma undergoes a phase transition to a hadron gas. At around the QCD phase transition temperature, we predict that for values of $\zeta/s\gtrapprox 0.1$, cavitation in the QGP will occur.
Under the assumption that experimental data on the QGP from \hic is inconsistent with the presence of hadron gas bubbles, our results for $\bulktext$ may be interpreted as an upper bound on the bulk viscosity in high-temperature QCD.  At very high temperatures, this interpretation seems consistent with known perturbative values of $\zeta/s$.
Several aspects of our work can and should be improved in subsequent studies: first, it is possible to implement the corrections from bulk viscosity in the flow profiles used in the calculations of $\bulktext$, eliminating the approximation we have used in this work. Second, one can repeat our study with more realistic approximations for the hadron gas pressure than our choice: $\pvap\equiv0$. 
Ultimately, and maybe most importantly, it would be interesting to calculate the particle spectra from a numerical simulation including the presence of hadron gas bubbles. This could potentially be done using state-of-the-art numerical hydrodynamical solvers \cite{RomatschkeFullyRelLatticeBoltzmannAlgorithm,SchenkeEventbyEventAnisotropicFlowHICYM,Denicol:2014vaa} and could verify the assumption that cavitating fluids are inconsistent with experimental data on \hic.

\section{Acknowledgements}
This work was supported by the \emph{Sloan Foundation}, Award No. BR2012-038
and the \emph{Deutsche Forschungsgemeinschaft}  (DFG), Grant No. RO 4513/1-1.

\bibliographystyle{JHEP}
\bibliography{references}
\end{document}